\documentclass[preprint]{imsart} 

\RequirePackage{amsthm,amsmath,amsfonts,amssymb}
\RequirePackage[authoryear]{natbib}
\usepackage{float,color}
\RequirePackage[colorlinks,citecolor=blue,urlcolor=blue]{hyperref}
\RequirePackage{graphicx}

\startlocaldefs

\newcommand{\bbeta}{\boldsymbol{\beta}}
\newcommand{\btau}{\boldsymbol{\tau}}

\newcommand{\btheta}{\boldsymbol{\theta}}

\newcommand{\bx}{ \mbox{\bf x}}
\newcommand{\bc}{ \mbox{\bf c}}
\newcommand{\beqn}{ \begin{eqnarray}}
\newcommand{\eeqn}{ \end{eqnarray}}

\newcommand{\Tr}{^{\rm T}}

\newcommand{\n}{{\bf n}}

\newcommand{\ben}{\begin{enumerate}}
\newcommand{\een}{\end{enumerate}}
\newcommand{\beq}{\begin{equation}}
\newcommand{\eeq}{\end{equation}}
\newcommand{\bde}{\begin{description}}
\newcommand{\ede}{\end{description}}

\newcommand{\norm}[1]{\lVert#1\rVert}

\newcommand{\bX}{{\bf X}}

\newcommand{\iid}{\stackrel{\mathrm{iid}}{\sim}}

\newcommand{\kl}[2]{\mathrm{KL}\mbox{$\left( #1 \mid \mid #2 \right)$}}
\newcommand{\skl}[2]{d_{\mathrm{KL}}\mbox{$\left( #1 \mid \mid #2 \right)$}}
\newcommand{\Reals}[1]{\mathbf{R}^{#1}}
\newcommand{\Real}{\mathbf{R}}

\newcommand{\domain}{\mathcal{D}}
\newcommand{\task}{\mathcal{T}}
\newcommand{\tildetau}{\tilde{\tau}}
\newcommand{\tildealpha}{\tilde{\alpha}}

\newtheorem{theorem}{Theorem}

\newtheorem{definition}[theorem]{Definition}

\endlocaldefs

\begin{document}

\begin{frontmatter}
\title{A penalized complexity prior for deep Bayesian transfer learning with application to materials informatics}
\runtitle{Bayesian transfer learning}

\begin{aug}
\author[]{\fnms{Mohamed A} \snm{Abba}\ead[label=e1,mark]{mabba@ncsu.edu}},
\author[]{\fnms{Jonathan P} \snm{Williams}\ead[label=e2,mark]{jwilli27@ncsu.edu}}
\and
\author[]{\fnms{Brian J} \snm{Reich}\ead[label=e3,mark]{bjreich@ncsu.edu}}
\address[]{North Carolina State University, \printead{e1,e2,e3}}
\end{aug}

\begin{abstract}
A key task in the emerging field of materials informatics is to use machine learning to predict a material's properties and functions.  A fast and accurate predictive model allows researchers to more efficiently identify or construct a material with desirable properties.  As in many fields, deep learning is one of the state-of-the art approaches, but fully training a deep learning model is not always feasible in materials informatics due to limitations on data availability, computational resources, and time.  Accordingly, there is a critical need in the application of deep learning to materials informatics problems to develop efficient {\em transfer learning} algorithms.  The Bayesian framework is natural for transfer learning because the model trained from the source data can be encoded in the prior distribution for the target task of interest.  However, the Bayesian perspective on transfer learning is relatively unaccounted for in the literature, and is complicated for deep learning because the parameter space is large and the interpretations of individual parameters are unclear.  Therefore, rather than subjective prior distributions for individual parameters, we propose a new Bayesian transfer learning approach based on the penalized complexity prior on the Kullback–Leibler divergence between the predictive models of the source and target tasks.  We show via simulations that the proposed method outperforms other transfer learning methods across a variety of settings.  The new method is then applied to a predictive materials science problem where we show improved precision for estimating the band gap of a material based on its structural properties.
\end{abstract}


\end{frontmatter}

\section{Introduction}\label{s:intro}

Materials informatics has fundamentally changed materials science research \citep[e.g.,][]{himanen2019data}.  To design or select a material for a particular function, researchers have traditionally relied on intuition and costly experimentation.  This process is now supplemented by machine learning to predict a candidate material's properties and triage materials for further experimentation. The addition of machine learning has been shown to improve efficiency, especially when using multiple data sources \citep[e.g.,][]{batra2021accurate}.  We  aim to build a predictive model for a material's band gap, defined as the energy differential between the lowest unoccupied and highest occupied electronic states \citep{kittel2019sons}.  The band gap governs desirable properties that are useful in industrial sectors such as electric and photovoltaic conductivity.  Traditionally, band gap size is computed using methods in quantum mechanics such as density functional theory \citep[DFT; ][]{DFT_citation}.  However, these methods require running costly computer simulations, and so it is {\em not} feasible to exhaustively search over a broad class of materials. Our objective is to build a statistical model \citep[sometimes called a meta-model, surrogate model, or emulator;][]{o2006bayesian} along with an accompanying transfer learning methodology that is both effective at predicting the output of the DFT simulation, and is able to optimize experimentation on future test data sets from related data sources.

Deep neural network (DNN) architectures have emerged as leading models in materials informatics.  Beyond materials science, DNNs have revolutionized the field of machine learning with significant breakthroughs in a wide variety of applications including computer vision \citep{comp_vision_voulodimos2018deep}, natural language processing \citep{nlp_young2018recent}, and protein folding \citep{protein_senior2020improved}; see \cite{dargan2019survey} for further references to modern applications. Their ability to model complex nonlinear processes enables them to handle a large class of prediction problems.  Training a neural network, however, is not an easy task, often requiring intensive computational cost, large quantities of training data, and careful hyperparameter selection \citep{bengio2012practical_parameter_selection}.  As an over-parameterized expressive model, DNNs are prone to overfitting, especially in the case of small data sets.  \cite{tan2018survey_dtl} concludes that the number of parameters in a DNN and the size of the data required for good generalization performance have an almost linear relationship. Since these models tend to have thousands (sometimes millions) of parameters, the size of the data required becomes quickly prohibitive. This is the where \textit{transfer learning} plays a central role.

Broadly defined, transfer learning describes a machine learning approach for augmenting the training of a learning task on a {\em target} population data set with a learning algorithm that has already been trained on a closely related data set from a {\em source} population, particularly for scenarios where the target and source populations are {\em not} identical; see \cite{weiss2016survey} for a recent survey on transfer learning.  By developing effective transfer learning strategies, it is possible to reduce the computational burden, and the need for large training data sets, for training DNN models in applications where complex DNNs have already been trained for similar data sets.  For example, \cite{dl_medical2021} and 
\cite{DBLP:journals/corr/abs-1902-07208} investigated the use of a pre-trained DNN on ImageNet \citep{deng2009imagenet} to improve the accuracy in medical imaging tasks where labeled data is usually scarce. Although the potential benefits for transfer learning are significant, there are no generally accepted procedures for how to construct or evaluate a transfer learning strategy, and this is especially true from the Bayesian perspective.  \cite{dube2020improving} provides a review of the current approaches to address a variety of important questions of concern for developing transfer learning methods.  For instance, what aspects of the source tasks can be leveraged to improve performance on the target task?  In the case of DNN source models, this commonly boils down to isolating the layers that are task agnostic versus those that are task dependent.   

Motivated by the predictive materials science application of estimating the band gap of a material based on its structural properties, we propose a principled statistical approach to transfer learning within a Bayesian framework.  Bayesian neural networks have attracted considerable attention in recent years as a result of computational advances \citep[e.g.,][]{wilson2020case_bnn_advances}.  However, the only work (that we are currently aware of) on deep Bayesian transfer learning is \cite{Wohlert2018BayesianTL}, where a single source task is considered and a DNN with feed forward layers is trained using mean field variational Bayes \citep[VB;][]{zhang2018advances_mfvb}. The approximate posterior learned on the source task is used as a prior on the parameters for the target task. \cite{Wohlert2018BayesianTL} did not, however, address the problem of freezing or training the transferred layers, and surprisingly gave worse performance than training using only the target data.  More generally, Bayesian transfer learning for models other than DNNs has been considered by other articles in the literature, mostly likely attributable to the fact that it is natural to borrow strength across data sets via prior distributions \citep[e.g.,][]{karbalayghareh2018optimal_bayesian_tl}.  

The Bayesian transfer learning methodology that we develop sequentially trains a DNN on a target data set by first specifying a prior distribution concentrated on parameter estimates for the DNN trained from a source data set.  While this approach is sensible, the primary challenge is that constructing prior densities from pre-trained DNN layers is not trivial because individual parameters are not identifiable nor do they have inherent meaning.  Accordingly, to solve this issue, we extend the fundamental notion of {\em penalized complexity priors} (PCPs) from \cite{simpson2017penalising} to the deep learning setting, and specify our {\em transfer learning prior distribution} in terms of the Kullback-Leibler divergence between the source and target model.  This prior is constructed with the assumption that the predictive model is likely shared (to some extent) across tasks, but is flexible enough to disregard the source task if appropriate.

Since the target data set in our materials science application is small, we implement fully Bayesian analysis with computations via Markov chain Monte Carlo (MCMC) sampling strategies.  However, since the source data set is large, a fully Bayesian analysis is prohibitive, and so we analyze the source data both using optimization and variational inference (VI) \citep{blundell2015weight} methods.  Both methods provide point estimates for each parameter, but VI also provides (approximate) uncertainty quantification.  Therefore, in addition to applying to cases where data sets from both sources are available, our proposed method can be applied using source data DNNs trained by other users, such as ImageNet \citep{deng2009imagenet}.  In our exposition, we develop our transfer learning methods using the optimization-based source model and the VI-based source model, in parallel, and we show that both provide improvements over traditional transfer learning methods for both synthetic and real data.

The remainder of the paper proceeds as follows.  Section \ref{s:background} reviews background material on DNNs and transfer learning.  Section \ref{s:model2} introduces the proposed PCP priors for transfer learning in DNNs.  The proposed methods are then evaluated using simulated and real data in Sections \ref{s:sim} and \ref{s:analysis}, respectively.  Section \ref{s:conclusions} concludes. The code for reproducing our empirical results is available at \url{https://github.ncsu.edu/mabba/Pybayes.git}. 

\section{Background material}\label{s:background}

\subsection{Deep Learning Model}\label{section: DL}
We begin by briefly reviewing the DNN model; for a more comprehesive review see \cite{goodfellow2016deep}.  In general, DNNs are comprised of successive layers, starting with an input layer, followed by one or more hidden layers, and finally an output layer. Each layer contains a number of nodes that are connected to the next layer through weights that control the impact of a given layer on the next.  For concreteness, we describe the Bayesian transfer learning model in the simple case of a continuous response and fully-connected feed forward neural network.  However, the ideas proposed in this paper translate to other structures, including recurrent \citep{medsker2001recurrent} and convolutional \citep{lecun1995convolutional_nets} networks (see Section \ref{s:conclusions} for further discussion).


A dense feed forward layer is a function $l: \Reals{d} \longrightarrow \Reals{k}$ characterized by a weight matrix $W \in \Reals{k \times d}$, a bias vector $b \in \Reals{k}$, and a real-valued activation function $a$ applied element wise to a vector argument. For $\bx \in \Reals{d}$, the dense layer applies an affine transformation to $\bx$ followed by the activation function, $l(\bx)  = a(W\bx+b)$.  A DNN with $L$ hidden layers first applies an input layer, typically $l_0(\bx) = \bx$, then $L-1$ hidden layers, and one output layer; each layer is assigned unique weights, biases, and (potentially) activation functions, i.e., $l_i(\bx)=a_i(W_i\bx+b_i)$ for layer $i\in\{1,...,L\}$.  The composition of these $L$ layers defines a function $f$ from the input space $\mathcal{X}$ to the output space $\mathcal{Y}$:
\begin{align}\label{eq: DNN}
    f(\cdot; \btheta) & : \mathcal{X} \longrightarrow \mathcal{Y}  \nonumber \\
    f(\bx; \btheta) & = l_L(l_{L-1}(\cdots (l_1(\bx)))),
\end{align}
where $\btheta :=\{\btheta_1,...,\btheta_L\}$ and $\btheta_i := \{b_i,W_i\}$ are the parameters for layer $i$.
For the regression tasks, we consider $f(\bx; \btheta)$ 
to be the mean response given $\bx$, whereas for a classification task, $f(\bx; \btheta)$ represents a vector of probabilities.

\subsection{Transfer Learning}\label{section: TL}

Before discussing transfer learning for DNNs, we review general definitions and concepts of transfer learning. We adopt notation from \cite{pan2009survey} and \cite{weiss2016survey} whose early work has been widely adopted in the literature.  A {\em domain} is composed of two parts, a feature space $\mathcal{X}$ and a sampling distribution $p(\bX)$, where $\bX$ denotes a set of instances $\bX = \left\{ \bx_i \mid \bx_i \in \mathcal{X}, i=1, \ldots,n \right\}$.  Accordingly, $\domain = \left\{ \mathcal{X}, p\right\}$.  Next, a {\em task} consists of a label space $\mathcal{Y}$ and a decision function $f$, that is, $\task = \left\{ \mathcal{Y}, f \right\}$.
For a {\em source} domain and task, denoted $\domain_S$ and $\task_S$, respectively, an observed data set is the collection $\left\{ (\bx_i, y_i) \mid \bx_i \in \mathcal{X}_S, y_i \in \mathcal{Y}_S, 1 \leq i \leq n_S \right\}$, and $f_{S}$ is understood as a conditional distribution or decision function for the instances $y_{i}$ given $\bx_i$ for $i \in \{1,\dots,n_{S}\}$.

\begin{definition}\label{def: TL}(Transfer Learning) Given $t$ data sets corresponding to source domains and tasks $(\domain_{S_1}, \task_{S_1}), \dots, (\domain_{S_t}, \task_{S_t})$, along with a data set from a target domain and task $(\domain_T, \task_T)$, transfer learning describes any strategy of using learned/trained conditional distribution or decision functions $f_{S_{1}}, \dots, f_{S_{t}}$ to augment the learning/training of the conditional distribution or decision functions $f_T$ on the target domain.
\end{definition}

In Definition \ref{def: TL}, for $t>1$ describes multi-source transfer learning  \citep[e.g.][]{maurer2016benefit}. Based on the assumptions about the relationships between the various domains and tasks, transfer learning is further classified into a variety of categories \citep{pan2009survey}.  DNNs typically require \textit{inductive transfer learning} where the feature space is the same across tasks, i.e., $\mathcal{X}_S = \mathcal{X}_T$.  Therefore, we will drop the subscript on the feature space and simply use $\mathcal{X}$. Note that this does {\em not} imply that the conditional decision functions, nor the sampling distributions of the features, are the same for the source and target.

\subsection{Transfer Learning with DNN architectures}

\begin{figure}[H]
    \centering
    \includegraphics[width = 0.7\linewidth]{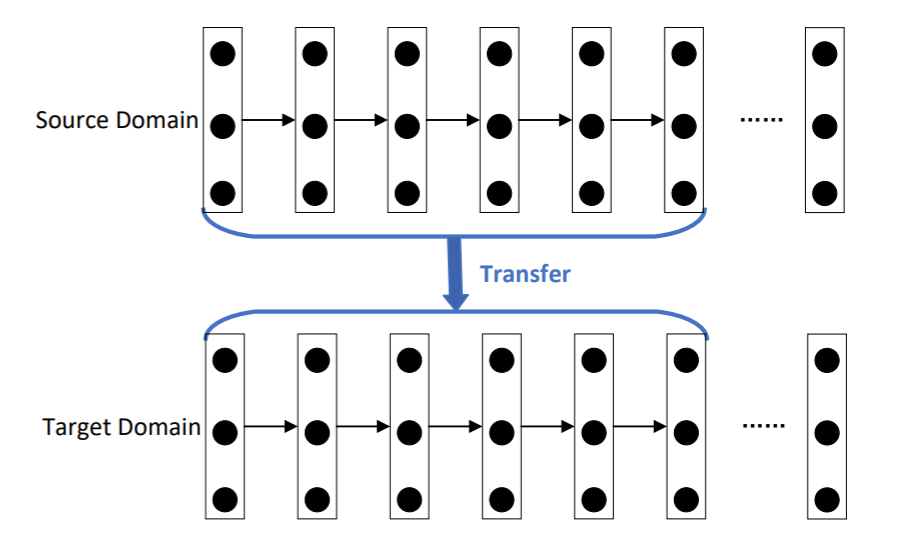}
    \caption{Sketch of network transfer; figure reproduced from \cite{tan2018survey_dtl}.}
    \label{fig: chaunqui_fg}
\end{figure}

As discussed in Section \ref{section: DL}, DNNs are composed of many layers, each performing a transformation of the data.  These transformations can be viewed as learned representations. It has been argued that different layers may learn different concepts relating to the task \citep{dube2020improving}. Specifically, there is evidence that early layers learn general representations of the data, while the the deeper layers are more task specific and learn specialized representations. In \cite{DBLP:journals/corr/ZeilerF13}, the authors studied the activation of early layers of DNNs in computer vision problems, where convolutional layers are typically used, and noted that early convolutional layers extract high level features with the potential to generalize to different image domains. Using this logic, it follows that transfer learning for DNNs amounts to determining how many early layers to share across tasks and how many to assign as task specific. 

The process of inductive transfer learning in DNNs can be summarized as:
\begin{itemize}
 \item Let $g_{\btheta}: \mathcal{X} \longrightarrow \Reals{k}$ be the function parameterized by $\btheta$ representing the shared architecture across all tasks, and let $\btheta = \btheta_S$ on the source tasks and $\btheta = \btheta_T$ for the target task.
 \item Let $f^i_{\bbeta_{i}}: \Reals{k} \longrightarrow \mathcal{Y}_{\mathcal{S}_i}$, for $i \in \{1,\dots,t\}$, be activation functions parameterized by some $\bbeta_{i}$ and representing the top layers for each source task.
 \item Let $f^0_{\bbeta}: \Reals{k} \longrightarrow \mathcal{Y}_{\mathcal{T}}$ be the top layer activation function of the model for the target task, with $\bbeta_{0}$ its corresponding parameters.
\end{itemize}
Accordingly, a transfer learning strategy with DNNs can be defined as the process of learning simultaneously the parameters $\btheta_S, \bbeta_{1}, \dots, \bbeta_{t}$ on the source tasks, and subsequently using the learned feature representation parameters for $\btheta_S$ to augment the learning of $\btheta_T$ on the target task. The transferred layers can either be held fixed on the target task (referred to as {\em freezing}) or fine-tuned. Recently, \cite{dube2020improving} investigated the empirical performances of freezing and fine-tuning. In their work, they focused on a computer vision task with pre-trained layers from different architectures that were fit to the Imagenet data set. The conclusion was that freezing pre-trained layers, although not optimal, was always better that random initialization. The best performance was obtained when the learning rate for the pre-trained layers was significantly lower than the rate of the other layers in the model. They advocate for a factor of $10\%$ as a default choice.

\section{PCPs for Deep Bayesian Transfer Learning}\label{s:model2}

Given $t$ source tasks and a target task as in Definition \ref{def: TL}, for the target task let $\btheta_T \equiv\btheta= \{\btheta_1,...,\btheta_{L}\}$ be the parameters associated with the architecture shared with the source tasks, denote $v_i$ as the number of parameters in $\btheta_i$ and take $\bbeta=\bbeta_T$ be parameters specific to the target task.  Denote $\hat{\btheta} = \{{\hat \btheta}_1,...,{\hat \btheta}_{L}\}$ as the estimated values for $\btheta$ that are trained from the source data and, when available, let $\Sigma_i$ be the covariance matrix for $\btheta_i$ (e.g., the posterior covariance given the source data).  Our goal is to leverage these estimated parameters to construct an informative prior for the target task. If the tasks were the same the natural solution would be to set $\btheta = \hat{\btheta}$. On the other hand, if the tasks were not at all related, $\hat{\btheta}$ does not encode any information for the target task and would probably be worse than random initialisation. In between these two extreme cases lies the core of transfer learning, where there is a reason to assume that tasks are related but we do not know to which degree.  

The focus of our analysis is a judicious choice of the prior distribution for $\btheta$. Constructing a prior distribution that uses ${\hat \btheta}$ requires, first, the choice of a method for computing ${\hat \btheta}$ from the source data, and second, requires specification of the prior distribution for $\btheta$ given ${\hat \btheta}$.  As described in the next subsection, we use the PCP of \cite{simpson2017penalising} for the prior distribution of $\btheta$ given ${\hat \btheta}$.  The remainder of this section provides an overview the PCP idea of \cite{simpson2017penalising}, and then describes two methods we propose for constructing ${\hat \btheta}$.  The PCP is derived for both methods of constructing ${\hat \btheta}$.

\subsection{Penalized Complexity Priors}

In the framework of \cite{simpson2017penalising}, the problem of prior choice is considered from a model complexity perspective. The prior on the parameter of interest $\theta$ is controlled by a flexibility parameter $\tau$. The base (or simplest) model for $\theta$, that we will denote by $p_{0}(\theta)$  corresponds to the case where $\tau=0$, and for larger values of $\tau$ the prior deviates from the base model. Following Occam's razor, simple models should be preferred until there is evidence for more complex ones; in other words, the hyperprior on $\tau$ should favor the base model. PCPs as defined in \cite{simpson2017penalising} give a principled way of choosing a prior on $\tau$ that controls the deviation from the base model by penalizing a scaled version of  the Kullback-Liebler divergence from $p_{0}(\theta)$ to $p(\theta | \tau)$:
\begin{equation}\label{eq: scaled_kl_div}
 \skl{p}{p_{0}} = \sqrt{2\kl{p}{p_0}} = \left(2\int_{\theta} p(\theta | \tau ) \log \left( \frac{ p(\theta | \tau)}{ p_{0}(\theta)}  \right) d \theta \right)^{1/2}.
\end{equation}
From \eqref{eq: scaled_kl_div}, it is clear that $\skl{p}{p_0}$ is a function of $\tau$. That being so, let $$ h(\tau) = \skl{p(\theta | \tau)}{p_0(\theta)},$$ and notice that  $h(\tau) = 0 \mbox{ if and only if } \tau = 0$. Furthermore, if $h(\cdot)$ is strictly increasing, penalising $h(\tau)$ for large values induces a prior that penalizes large values of $\tau$, and consequently puts more mass close to the base model. \cite{simpson2017penalising} recommend using an exponential distribution, $h(\tau)\sim\mbox{Exp}(\lambda)$, to ensure a mode at $\tau=0$ and a constant rate of penalization for larger $\tau$.

For transfer learning, we place an uninformative prior on $\bbeta\sim\mbox{Normal}({\hat \bbeta}_0,\tau_0\bf{I})$  (e.g., ${\bbeta_0}=0$ and large $\tau_0$) and define
$$
  p(\btheta|\btau) = \prod_{i=1}^LN\left(\btheta_i; {\hat \btheta}_i,g(\tau_i)\Sigma_i\right),
$$
where $N$ is the multivariate Gaussian density function, $g$ is an increasing function, the flexibility parameter $\tau_i$ controls the deviation from the base model in layer $i$ and $\btau=(\tau_1,...,\tau_L)$. The case $\btau = \mathbf{0}$ is used to define the base model. This layer-wise specification reflects the fact that some layers can be more transferable than others, and hence require more penalization. 

If we marginalize over $\btau$, the prior model becomes a scale mixture of Gaussian distributions, where the mixing distribution will be the prior on $\btau$. This latter distribution will be implicitly defined by placing an exponential distribution on $h(\btau) = \skl{p(\btheta_T | \btau)}{p_0(\btheta_T)}$. When $L > 1$, the implicit prior on $\btau$ is not identifiable since the mapping $h: \Reals{L}_{+} \longrightarrow \Real_{+}$ cannot be bijective and hence is not invertible. To circumvent this issue, 
\cite{simpson2017penalising} extend the PCP prior to a multivariate parameter $\btau$  by having the prior on $\btau$ uniform on each level set of $h(\cdot)$. Instead, we opt for the simpler approach of  constraining $\btau$ to the $(L-1)$-simplex given by any weighted combination of its elements.  This is equivalent to $\btau = \tilde{\tau}\bc$, where $\bc \sim \mbox{Dirichlet}(\mathbf{1}_L)$ and $\tilde{\tau}\in \Real_{+}$. In this formulation, $\tilde{\tau}$ becomes a global flexibility parameter, and the weights $c_i$ control how much deviation each layer is allowed since $\tau_i = \tilde{\tau}c_i$. Thus, given the vector of weights $\bc$, we have a one-to-one mapping $h_{\bc}(\tilde{\tau})$, through which we have an explicit penalising prior on the global parameter $\tilde{\tau}$.

\subsection{PCP for the VI case}\label{s:VI}
If the source data are analyzed using VB methods, then ${\hat \btheta}_i$ is an approximate posterior mean and $\Sigma_i$ is an approximate $v_i\times v_i$ posterior covariance matrix.  Because standard VB methods approximate the posterior as independent across parameters, $\Sigma_i$ is a diagonal matrix for each $i$.  We incorporate this information in the prior by setting $g(\tau) = 1+\tau$ so that the base model with $\btau=\mathbf{0}$ uses the posterior of the source data directly as the prior for the target data, which would be optimal Bayesian learning if the two tasks are the same.  If $\tau_i>0$, then the prior variance increases to reflect uncertainty about the relationship between source and target tasks.

Under this prior distribution, the KL divergence is
\begin{align*}
    \kl{p(\btheta | \tilde{\tau}, \bc)}{p_{0}(\btheta)} = & \frac{1}{2}\left( \sum_{i=1}^L v_i c_i\tildetau  - v_i\log \left(1+c_i\tildetau   \right) \right).
\end{align*}
So if $d = \sqrt{2\kl{p(\btheta | \tilde{\tau}, \bc)}{p_0(\btheta)}} \sim \mbox{Exp}(\lambda)$, the one-to-one mapping between $d$  and $\tildetau$ is
\beq\label{tau-dist: VI mapping}
d = \sqrt{\sum_{i=1}^L v_i \phi(c_i\tildetau)}\eeq
where $\phi(x) = x - \log(1+x)$. Furthermore, $\phi$ is continuously differentiable and strictly increasing, hence given $\bc$ we have a unique prior on $\tildetau$ corresponding to the desired prior on the KL divergence from the base model.  The induced prior is
\begin{align}\label{eq: vi_tau_prior}
 p(\tildetau | \bc) = & p(h_{\bc}(\tildetau))  \left| \frac{dh_{\bc}(\tildetau)}{d\tildetau} \right| =   \frac{\lambda e^{-\lambda h_{\bc}(\tildetau)}}{2\left( \sum_{i=1}^{L} v_i \phi(c_i\tildetau)  \right)^{1/2}} \left( \sum_{i=1}^{L} v_i \frac{c_i^2 \tildetau}{1 + c_i\tildetau}  \right).
\end{align}
with $h_{\bc}(\tildetau)^2 =  \sum_{i=1}^{L} v_i \phi(c_i\tildetau)$.  Equation \eqref{eq: vi_tau_prior} coupled with the Dirichlet prior on the weight $\bc$ completely specifies a prior distribution on the vector of scales $\btau$. Using numerical inversion of the mapping in \eqref{tau-dist: VI mapping}, sampling from this prior is straightforward.
\begin{figure}[H]
    \centering
    \includegraphics[width = 0.5\textwidth]{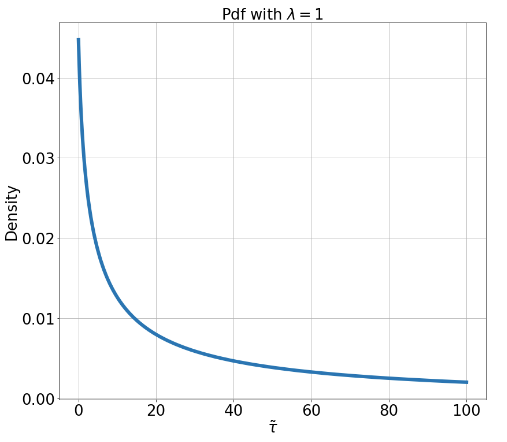}
    \caption{The form of the PCP distribution on $\tildetau$ with $\lambda=1$, $L = 3, c_i=1/L$, $v_1=v_2=16$ and $v_3 = 1$.}
    \label{fig: tau_pdf}
\end{figure}

It can be shown that the induced prior on the global flexibility parameter in (\ref{eq: vi_tau_prior}) has a mode at zero with a strictly decreasing density as $\tildetau$ increases. Also, the tail of the prior behaves like a modified Weibull distribution \citep{almalki2014modifications} with rate $\lambda \sum v_ic_i$ and shape $0.5$. 
Figure \ref{fig: tau_pdf} plots the density when $\lambda = 1$. 

\subsection{PCP for the point estimate case}\label{s:point_estimate}
Assume that a point estimate ${\hat \btheta}$ is derived from the source data but no measure of uncertainty $\Sigma$ is provided.  In this case, the source data does not define a base distribution for $\btheta$ and thus we cannot compute the KL divergence between base and full models for $\btheta$ without further assumptions.  Since DNNs are primarily used for prediction, we place a prior on the KL divergence between the predictive model under the source and target models.   Let $f(\bx | \btheta, \bbeta)$ denote the target DNN output for a given input $\bx$ and parameter values $(\btheta,\bbeta)$.  The prior distributions $p(\bbeta) \sim \mbox{Normal}( \hat{\bbeta}, \tau_0\mathbf{I})$ and $\btheta_i\sim\mbox{Normal}({\hat \btheta}_i,\tau_i\bf{I})$ induce a prior distribution on the function $f$.  Unfortunately, for the DNN model it is not available in analytical form. To overcome this, we approximate the DNN output using a first-order Taylor approximation around $(\hat{\btheta}, \hat{\bbeta}) $, where $\hat{\bbeta}$ is an appropriate initialization of the top layers:
\begin{equation}\label{eq: first_order}
 f(\bx | \btheta,\bbeta)  \approx f(\bx | \hat{\btheta}, \hat{\bbeta}) + \left(\bbeta - \hat{\bbeta}\right)\Tr\nabla_{\hat \bbeta}f(\bx | \hat{\btheta}, \hat{\bbeta}) + \left(\btheta - \hat{\btheta}\right)\Tr\nabla_{\hat \btheta} f(\bx | \hat{\btheta}, \hat{\bbeta}).
\end{equation}
For small values of $\tildetau$ (i.e., the most relevant values for the prior, assuming the target and source are similar) a first-order approximation is arguably tight enough.  Furthermore, if the activation functions used in the model are piece-wise linear like the widely used ReLU  function \citep{agarap2018deep}, the output function $f(\bx | \btheta,\bbeta)$ will also be piece-wise linear and we cannot go beyond a first-order Taylor expansion.

Based on the approximation, for the base model with $\btau=\mathbf{0}$ we get the prior distribution 
$$ p_{0}\left(f(\bx)\right) \sim \mbox{Normal}\left( f(\bx | \hat{\btheta}, \hat{\bbeta}), \norm{\nabla_{\hat{\bbeta}}f(\bx | \hat{\btheta}, \hat{\bbeta})}^2  \right),$$
 and for the flexible model the prior on $f(\bx | \tildetau, \bc)$ is
$$ p\left( f(\bx)| \tildetau, \bc \right) \stackrel{.}{\approx} \mbox{Normal}\left( f(\bx | \hat{\btheta}, \hat{\bbeta}), \norm{\nabla_{\hat{\bbeta}}f(\bx | \hat{\btheta}, \hat{\bbeta})}^2  + \tildetau \bc \odot \alpha(\bx) \right),$$ where $\alpha(\bx) = \left(\norm{\nabla_{\hat \btheta_i}f(\bx | \hat{\btheta}, \hat{\bbeta})}^2\right)_{1\leq i \leq L}$. Now that we have two normal models, an analytic expression for the KL distance is
\begin{align}\label{kl-expression}
    \kl{ p\left( f(\bx)| \tildetau, \bc \right)}{p_{0}} 
             = & \frac{1}{2} \phi\left[ \tildetau\tildealpha_{\bc}(\bx)\right] \ \ \mbox{ where } \ \ \tildealpha_{\bc}(\bx) = \frac{\bc \odot \alpha(\bx)}{\norm{\nabla_{\hat{\bbeta}}f(\bx | \hat{\btheta}, \hat{\bbeta})}^2};
\end{align}
the details of the derivation of the expression are relegated to the Appendix.

In equation \eqref{kl-expression}, $\tildealpha(\bx)$ depends on the gradient of $f(\bx | \hat{\btheta},\hat{\bbeta})$ for a given input point $\bx$. Assuming that the inputs are identically distributed and we have $N$ target data points, the marginal KL divergence can be approximated by
\begin{align}\label{kl-multi}
    \kl{ p
    }{p_{0}}  \approx & \frac{1}{2N} \sum_{i=1}^{N} \phi\left[ \tildetau\tildealpha_{\bc}(\bx_i)\right]. 
\end{align}
Using \eqref{kl-multi} we have a one-to-one mapping between $d = \sqrt{2\kl{p}{p_0}}$ and the global flexibility parameter $\tildetau$. Hence, an exponential distribution on $d$ induces a prior on $\tildetau$, which can be obtained with the same simple change of variables as in \eqref{eq: vi_tau_prior}. Let $d_{\bc}(\tildetau) = \sum_{i=1}^{N} \phi\left[ \tildetau\tildealpha_{\bc}(\bx_i)\right]/N$.  Then

\begin{align}\label{eq: pe_tau_prior}
 p(\tildetau | \bc) = & p(d_{\bc}(\tildetau))  \left| \frac{d_{\bc}(\tildetau)}{d\tildetau} \right| =   \frac{\lambda e^{-\lambda d_{\bc}(\tildetau)}}{2\left(\frac{1}{N} \sum_{i=1}^{N}  \phi\left[ \tildetau\tildealpha_{\bc}(\bx_i)\right] \right)^{1/2}} \left( \frac{1}{N}\sum_{i=1}^{N} \frac{\tildealpha_{\bc}(\bx_i)^2 \tildetau}{1 + \tildealpha_{\bc}(\bx_i)\tildetau}  \right).
\end{align}
 The form of the prior in \eqref{eq: pe_tau_prior} is similar to \eqref{eq: vi_tau_prior}: in both cases the prior induced on $\tildetau$ has mode at zero, is a strictly decreasing density as $\tildetau$ increases, and has the same tail behavior.

\begin{figure}[H]
    \makebox[\linewidth]{
        \includegraphics[width=\textwidth]{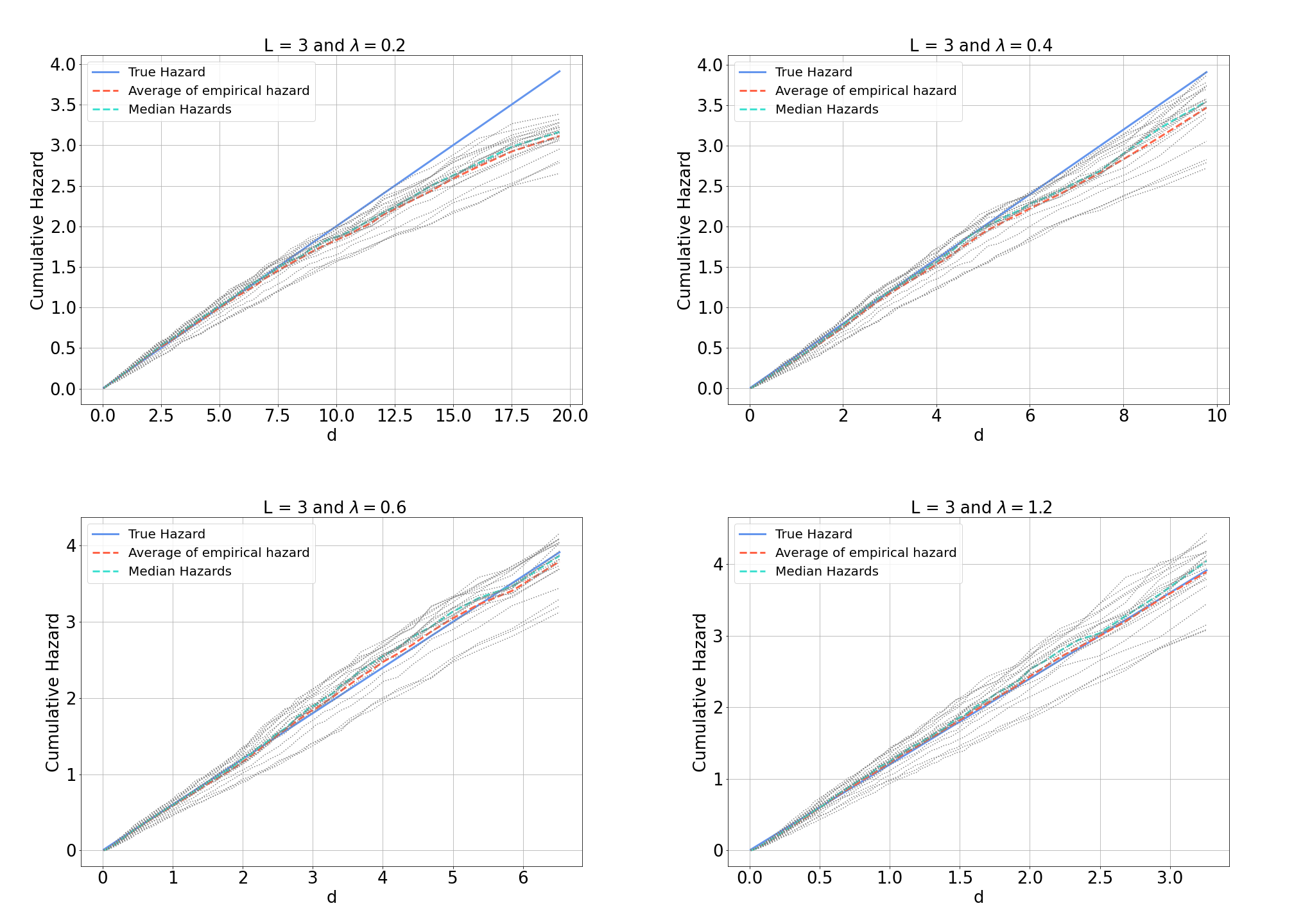}
    }
    \caption{ True hazard function of the scaled KL divergence (blue), and cumulative hazard for different generated data sets (gray). The dashed lines represent the median and mean of the empirical cumulative hazard.}
    \label{fig:kl_fit}
\end{figure}

We evaluate the approximation used in \eqref{eq: first_order} to verify that the derived prior on $\btau$ will result in an approximately exponential distribution on the \rm{KL} divergence between the two models.  We assume $L=3$ layers with $\{64, 32, 1\}$ nodes respectively, and omit the task-specific parameter $\bbeta$.  The source-data estimates ${\hat \btheta}_i$ are generated following the Glorot initialization scheme \citep{Glorot10understandingthe}.  Since we cannot compute the \rm{KL} divergence analytically, the distribution of KL divergence is approximated with Monte Carlo sampling. The approximation is evaluated for $50$ different target data sets generated as described in Section 4.  For each of those data sets we compute the gradients in \eqref{eq: first_order} and first draw $\bc$ and  ${\tilde \tau}$ from their priors for different values of the rate parameter $\lambda$, then sample $\btheta_i|\btau\sim\mbox{Normal}({\hat \btheta}_i,\tau_i\bf{I})$.  We then sample $20,000$ replications of the model output $f(\bx; \btheta)$ and approximate the KL divergence from the base model $f(\bx; \hat \btheta)$ for each replication  to get the empirical distribution of the KL divergence.  Figure \ref{fig:kl_fit} plots the cumulative hazard function of the scaled KL divergence $d$ which should be linear with slope $\lambda$ if the distribution is Exponential$(\lambda)$.  The approximation holds near the origin for all $\lambda$, and is tighter for the entire distribution for larger values of $\lambda$.

\section{Simulation study}\label{s:sim}

We perform the following simulation experiments to evaluate the predictive performance of the proposed Bayesian transfer learning methods.  The data are generated as follows.  For observation $i$, the covariates are generated as
$x_{ij}\iid\mbox{Normal}(0,1)$ and $x_{ij+1}|x_{ij}\sim\mbox{Normal}(\rho x_{ij},1-\rho^2)$ for $j>1$.  The source data are generated for constants, $c$, $\sigma$, and $k_{1}$, as
$$
Y_i\sim\mbox{Normal}\left\{c\mu_{1}(\bx_i),\sigma^2\right\} \ \ \text{with} \ \  \mu_{1}(\bx_i)=\cos\left(2\sum_{j=1}^{k_1}x_{ij}/k_1\right),
$$
and the target data are generated, for constant $k_{2}$, as 
$$
Y_i\sim\mbox{Normal}\left\{c\mu_2(\bx_i),\sigma^2\right\} \ \ \text{with} \ \ \mu_2(\bx_i)=\cos\left(2\sum_{j=1}^{k_2}x_{ij}/k_2\right).
$$
We set $p=30$ covariates, $\rho=0.5$, $\sigma=1$, $k_1 = 15$, and $k_2=k_1+k$ with $k\in\{0,5,10,15\}$, and we generate $1000$ synthetic source training observations, with a varying number of target training observations $n \in\{30, 50, 70\}$.  The constant $c$ is set so that the signal-to-noise ratio is 4:1 for the target data (i.e., so the variance of $c\mu(\bx_i)$ is 4).  If $k=0$ then the two data-generating processes are identical, and if $k$ is large they are dissimilar. For each scenario, we simulate 50 data sets and compare predictions on a test set of size $n_{\rm{test}}=200$.

\begin{table}[H]
\begingroup
\renewcommand{\arraystretch}{0.8}

\centering
(a) Mean squared prediction error (standard errors)\vspace{6pt}

\begin{tabular}{ll|ccc|ccc}
$k$ & $n$  & TL1 &   & TL2   & BNN   & BTL-PCP-PE & BTL-PCP-VI \\ \hline
&\vspace{-6pt}\\
0  & 30  &  1.94 (0.21) & & 1.33 (0.12) &  1.85 (0.19) &  1.09 (0.12) & 1.48 (0.16) \\
 0  & 50  &  1.76 (0.26) & & 1.17 (0.10) &  1.55 (0.10) &  1.02 (0.18) & 1.17 (0.34) \\     
 0  & 70 &  1.62 (0.22)& & 1.05  (0.15)&  1.08 (0.12)&  0.70 (0.04) & 0.93 (0.10) \\
 5 & 30  &  2.06 (0.23) & & 1.55 (0.17) &  1.84 (0.13) &  1.15 (0.09) & 1.53 (0.14) \\
 5  & 50  &  1.83 (0.15) & & 1.26 (0.09)&  1.62 (0.08) &  1.07 (0.11) & 1.21 (0.13) \\
 5 & 70 &  1.67 (0.17) & & 1.01  (0.20)&  1.03 (0.11) &  0.88 (0.08) & 0.96 (0.05) \\
 10  & 30  &  1.95 (0.35) & & 2.05 (0.18) &  1.91 (0.26) &  1.37 (0.10) & 1.78 (0.07)\\
 10 & 50  &  1.86 (0.19) & & 1.87 (0.27) &  1.64 (0.30) &  1.60 (0.19) &  1.94 (0.22) \\ 
 10 & 70 &  1.73 (0.31) & & 1.85 (0.16) &  1.05 (0.12) &  1.01 (0.14) & 1.16 (0.21) \\ 
 15  & 30  &  2.13 (0.24) & & 2.47 (0.09) &  1.88 (0.11) &  1.93 (0.07) & 2.27 (0.10) \\
 15 & 50  &  1.86 (0.28) & & 2.16 (0.14) &  1.49 (0.25) &  1.51 (0.20) &  1.90 (0.26)\\ 
 15 & 70 &  1.73 (0.19) & & 1.85 (0.20) &  1.22 (0.08) &  1.42 (0.04) & 1.76 (0.12)
\end{tabular} 
\vspace{12pt}

(b) Average coverage of 95\% prediction intervals\vspace{6pt}

\begin{tabular}{ll|ccc}
$k$ & $n$ & BNN & BTL-PCP-PE & BTL-PCP-VI \\ \hline
&\vspace{-6pt}\\
 0      &      30     & 0.88 & 0.93 & 0.77\\
 0      &      50     & 0.90 & 0.95 & 0.84\\
 0      &      70    & 0.92 & 0.99 & 0.89 \\
 5      &      30     & 0.86 & 0.89 & 0.73\\
 5      &      50     & 0.89 & 0.94 & 0.84\\
 5      &      70    & 0.91 & 0.97 & 0.90 \\
 10      &     30     & 0.84 & 0.88 & 0.76 \\
 10     &      50     & 0.90 & 0.94  & 0.91\\    
 10     &      70    & 0.95 & 0.96  & 0.90\\    
 15      &     30     & 0.79 & 0.77 & 0.63\\
 15     &      50     & 0.93 & 0.90  & 0.81\\ 
 15     &      70    & 0.96 & 0.95   & 0.87
\end{tabular}\caption{{\bf Simulation study results}: The non-Bayesian methods (``TL1-TL2'') use transfer learning by fixing different numbers of layers of the network using source data, the Bayesian methods are the Bayesian model without transfer learning (``BNN''), full Bayesian transfer learning model with point estimate (``BTL-PCP-PE'') and VI (``BTL-PCP-VI'') in the first stage.  Table (a) gives the mean square prediction error by the difference between the number of active covariates over tasks, $k$, and size of the target data set, $n$.  Table (b) gives the average coverage for the Bayesian 95\% prediction intervals.}
\label{t:sim1}
\endgroup\end{table}

Both the source and target data are analyzed using the fully-connected model with four hidden layers having widths 24, 16, 12, and 8, respectively.  The source data are fit using the optimization scheme given in Appendix A.2, which produces estimates ${\hat \btheta}$ and their variances, in the case of VI. We then fit several models to each target data set. The first group are non-Bayesian optimization-based transfer learning ``TL'' methods.  We consider two methods that fix some layers using the source fit and tune the remaining layers using the target data. The first model ``TL1'' shares no layers and is fit to the target data with no connection to the source data.  The second model ``TL2'' proceeds in two stages; first, the output layer only is trained on the target data while the others are frozen, and next, in the second stage all layers are fine tuned.  

The second group of methods are Bayesian. The first Bayesian model ``BNN'' ignores the source data by setting the prior $\btheta_i|\btau\sim\mbox{Normal}(\bf{0},\tau_i\bf{I})$ and $\tau_i\sim\mbox{InvGamma}(2,1)$ so that the marginal standard deviation is $1$.  The second method ``BTL-PCP-PE'' fits the Bayesian transfer learning model  with the priors centered at point estimates $\hat{\btheta}$ alone, as described in Section \ref{s:point_estimate}.  The third model ``Bayes-PCP-VI'' uses the PCP prior based on VI, as described in Section \ref{s:VI}.   For both PCP methods we assume $\bc\sim\mbox{Dirichlet}(\bf{1}_L)$ and set $\lambda$ so that the correlation between $f(\bx_i; \btheta)$ and $Y_i$ is $0.5$.  All Bayesian method are fit using MCMC as described in Appendix A.3.

The results are compared using mean squared error, and for the Bayesian methods the coverage of 95\% prediction intervals for $\mu_2(\bx_i)$ over a target data test set of 200 observations is provided.  The results are reported in Table \ref{t:sim1}. First, we see that Bayesian methods outperform the point estimate methods in all cases, perhaps due to the added stability of prior distributions. When the source and target are  similar (i.e., $k \in  \{0,5\}$), all the transfer learning methods work well, especially if the number of training samples is small.  When the discrepancy between the source and target domain is high, however, the transfer learning methods do not improve the performance of the models.   In fact, the point estimate method ``TL-1'' outperforms ``TL-2'', and the Bayesian methods are comparable. An interesting result is the poor performance of the PCP prior for the VI case compared to the other Bayesian methods. It seems that building a PCP prior based on the mean-field VI posterior approximation does not encode enough information for the model to learn from small data sets.

\section{Case Study: Molecular Gap Prediction}\label{s:analysis}

We apply the proposed transfer learning methods for band gap prediction.  Band gap size (measured in electronvolts, eV) is typically computed using different quantum mechanics methods like the Density Functional Theory \citep[DFT;][]{DFT_citation}. DFT calculation can be expensive for large molecules and crystals.  This, combined with the large possible number of materials, has driven the need for fast screening methods. According to \cite{band_gap_crystals}, there has been a growing interest in the development of accurate machine learning models for material sciences and quantum systems. For small organic molecules,  \cite{band_gap_crystals} reported that DNNs models have reached chemical accuracy, however for large molecular crystals the task of band gap prediction seems to be much harder. In addition, DFT calculations are more expensive to perform on molecular crystals. For example,  \cite{dft_time_complexity} found that the algorithmic complexity of DFT is cubic in the number of particles for a given molecule, hence it easier to collect and compute band gap values for molecules than it is for molecular crystals. Given the difficulty of band gap calculation for molecular crystals, we propose to use the transfer learning methods to borrow information from molecular crystals (source) to molecules (target) to improve prediction.

The target data considered here will consist of 50,000 molecular crystals and their band gap values computed using DFT. This data set was provided by the Material Science Research Group at Carnegie Mellon University and required a sustained effort over several years to produce. For training we consider only subsets of the data with cardinality 1,000, 2,500, or 5,000 samples to represent a typical data set size, a validation set of 10,000 samples and finally a testing set of 20,000 samples. The source data will be the OE62 data set \citep{mediatum1507656} which consists of 62,000 molecules and their band gap values.   The molecules in the source data were all extracted from organic crystals, and the band gap values were also computed using DFT. 

Since we are dealing with molecules, we first need to preprocess the data and compute descriptors $\bx$ that can be fed into the DNN. We use the MBTR descriptor method \citep{huo2018unified} to compute 1,260 descriptors for every data point in both the source and target data sets. These descriptors are functions of the structure of the material, e.g., the types and configuration of its atoms.  The architecture of the model will have two hidden layers with 128 and 64 neurons, respectively. First the model is trained on the source data with 40,000 samples for training set and 10,000 for the validation. The parameters of the best performing model in terms of mean squared error are saved and transferred to the target task as  $\hat{\btheta}$. 

On the target task, we consider four different transfer learning methods. The non-Bayesian methods differ in the treatment of the hidden layers of the model, we can either freeze the hidden layers and treat them as feature extractor and only fit the output layer to the target task ``TL-freeze'', or we can fine-tune the hidden layers along with the output layer on the target task ``TL-fine-tune''. For both Bayesian methods, the priors are $\btheta_i|\btau \sim \mbox{Normal}( \hat{\btheta}_{i}, \tau_i \mathbf{I})$. The two methods differ in the priors on $\btau=(\tau_1, \tau_2, \tau_3)$; we compare the PCP with point estimate approach ``BTL-PCP-PE'' in Section \ref{s:point_estimate} with $\bc\sim\mbox{Dirichlet}(1,1,1)$ and uninformative priors $\tau_i \sim \mbox{InvGamma}(2,1)$ ``BTL-BNN''.  Since the point-estimate approach outperformed the VI approach in the simulation study we focus on the point-estimate approach in this data analysis.  To verify that transfer learning provides benefit over using only the target data, we compare to the state of the art DNN model used in materials science where DNNs have been developed as end-to-end models in the sense that they take as input the molecular representation and learn their own embedding.  Two of the leading methods are SHNET \citep{schutt2017schnet} and MEGNET \citep{chen2019graph}.

\begin{figure}[H]
\centering
\includegraphics[width = 0.95\textwidth]{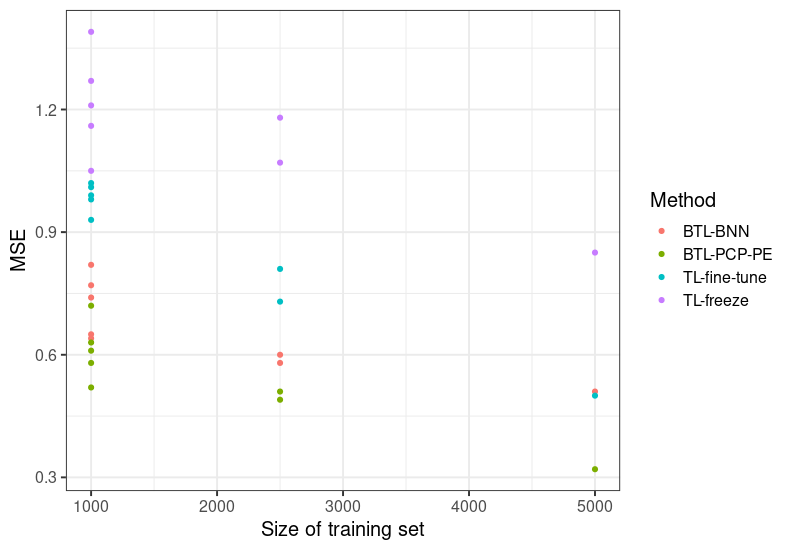}
\caption{Test set mean squared error (MSE; eV$^2$) for different transfer learning methods applied to band-gap prediction.  The methods are plotted against the size of the target data training set.  Each point represents the MSE for one fold in the cross validation.  For comparison, the sample variance of the response is $1.12$ eV$^2$.}
\label{fig:my_label}
\end{figure}

\begin{figure}[H]
    \centering
    \includegraphics[width=\textwidth]{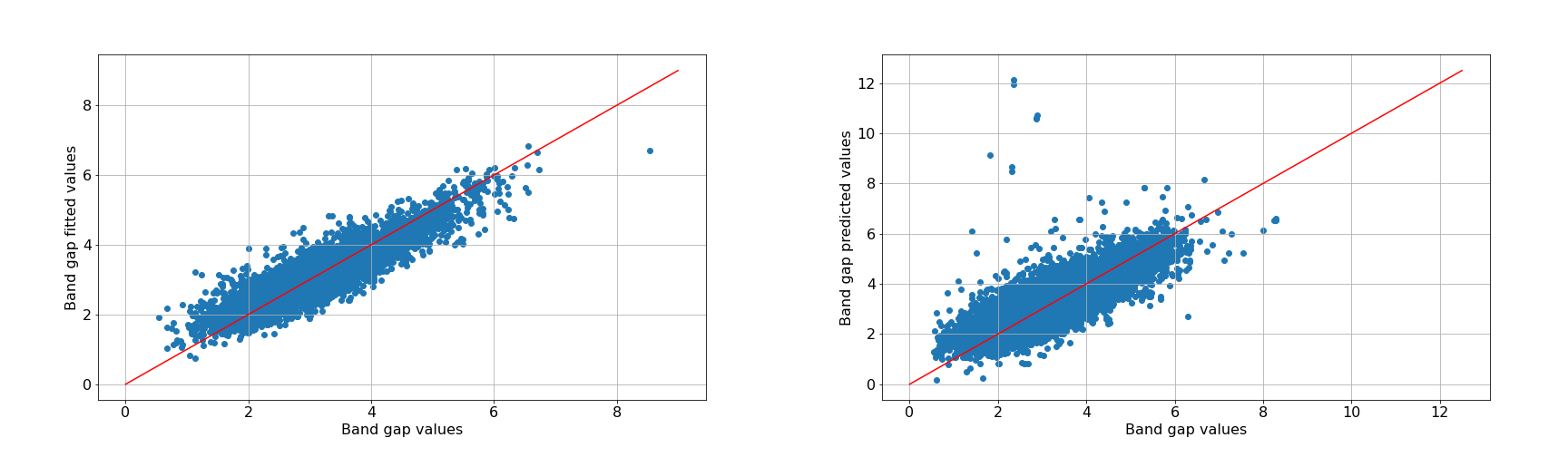}
    \caption{Predicted versus  band-gap for the training (left) and test (right) sets for the BLT-PCP-PE methods with $n=5,000$ training observations. The units of the plot are electronvolts.}
    \label{fig: prediction}
\end{figure}

The target data are split into groups of size 5,000, 10,000, and 20,000 for training, so that each group contains materials of different classes and the band gap distribution stays roughly the same.  We train the model in samples of size $n\in\{1000, 2500, 5000\}$, and Figure \ref{fig:my_label} plots the mean squared error (MSE) on a test set of size 20,000 for each method and training set, and Figure \ref{fig: prediction} plots the observed and predicted values for the BTL-PCP-PE method.  As expected, MSE decreases for each method as the size training set increases, and in all cases the BTL-PCP-PE method gives the smallest MSE.

With a full Bayesian treatment of all the parameters, we can look at the summary statistics of $\bc$, the parameter controls which layers  contain the useful information from the source data and which are irrelevant. The posterior means (standard deviation) for $c_1$, $c_2$ and $c_3$ are 0.05 (0.02), 0.05 (0.03) and 0.90 (0.15) respectively.  Therefore, the first two layers are shrunk towards the source data fits while the output layer varies from the source-model fit.


These results are competitive with state-of-the-art methods in the materials community.  For a training set of size 5,000 the MSE for SHNET and MEGNET are $0.48$ eV$^2$ and $ 1.01$ eV$^2$, respectively.  In fact, using the same data except with a much larger training set of 30,000 observations, the methods achieved MSE of 0.59 eV$^2$ for SHNET and 0.34 eV$^2$ for MEGNET.



\section{Conclusions}\label{s:conclusions}

Deep learning often provides excellent predictions when trained on large data sets, but is overparameterized and unstable for small data sets.  In this paper, we develop new transfer learning methods that provides regularization by centering the prior distribution on estimates from an auxiliary source data set.  We use the PCP to ensure that the prior concentrates around the source model but is allowed to deviate if appropriate.  We develop two methods for cases where the source data analysis does and does not provide uncertainty measures for the parameter estimates.  We show via simulation that the proposed methods reduce prediction error compared to standard transfer learning methods, and unlike standard methods the Bayesian approach gives reasonable coverage for prediction intervals.  The proposed methods are applied to band gap prediction where we show that transfer learning provides state-of-the-art accuracy.

There are many areas of future work.  We have developed our method in the simplest case of a continuous response and feed forward network. The prior based on VI in Section \ref{s:VI} would apply directly to other networks and response distributions as the prior is only a function of the approximate Gaussian posterior distribution from the source data analysis. Similarly, the prior based on point estimates in Section \ref{s:point_estimate} can be applied without modification to non-Gaussian responses and richer architecture.  Changing the network architecture would change the form of the gradients in (\ref{eq: first_order}) but the method itself can be used without modification.  The method also applies for non-continuous responses if we view $f$ as a process that spans the real line that is related to the response distribution.  For example, for binary data, $f$ in (\ref{eq: first_order}) could be the logistic function of the success probability, which spans the real line and thus the proposed methods could be applied on this scale.  

Next, our method applies an informative prior on the overall variance parameter, but a uniform prior on the distribution of the variance across levels, $\bc$. If prior information about the transferability of different layers is available, it could be incorporated in the prior for $\bc$. For example, it may be reasonable to assume the prior mean of $c_i$ increases from the input to the output layer. Lastly, another interesting area of future work is to compare the two-stage analysis that sequentially analyzes the source then target data with a simultaneous analysis of the all data sources.  Here we have assumed that building a hierarchical model for all data sources is computationally prohibitive, but it would be interesting to compare the efficiency of these two approaches when possible, especially when the source data set is small and thus the parameter estimates are uncertain.

\section*{Acknowledgements}

The authors thank the National Science Foundation (CMMT-1844484, CMMT-2022254, DGE-1633587), King Abdullah University of Science and Technology (3800.20), and the National Institutes of Health (1R56HL155373-01) for supporting this research. The authors also thank Noa Marom and Xingyu Liu of Carnegie Mellon University for assistance with the data collection and interpretation.

\section*{Appendix A.1: Derivations}\label{s:A1}

\subsection*{Approximation of the NN output}\label{Approx}

Under the base model, $\btheta = \hat{\btheta}$, and the last term in (\ref{eq: first_order}) cancels, on the other hand for the flexible model the prior is $\btheta_i\sim\mbox{Normal}({\hat \btheta}_i,\tau_i\bf{I})$, and the last term depends on the scale parameter $\tildetau$.  The prior on the top layers in the target task, using \eqref{eq: first_order}, we get the following distribution under the base model:
$$ p_{0}\left(f(\bx)\right) \stackrel{.}{\approx} \mathcal{N}\left( f(\bx | \hat{\btheta}, \hat{\bbeta}_{0}), \norm{\nabla_{\hat{\bbeta_{0}}}f(\bx | \hat{\btheta}_S, \hat{\bbeta}_{0})}^2  \right) ,$$
for the flexible model the prior on $f(\bx | \tildetau, \bc)$ becomes:
$$ p\left( f(\bx)| \tildetau, \bc \right) \stackrel{.}{\approx} \mathcal{N}\left( f(\bx | \hat{\btheta}_S, \hat{\bbeta}_{0}), \norm{\nabla_{\hat{\bbeta_{0}}}f(\bx | \hat{\btheta}_S, \hat{\bbeta}_{0})}^2  + \tildetau \bc \odot \alpha(\bx) \right),$$ $$\mbox{ where }  \alpha(\bx) = \left(\norm{\nabla_{\hat \btheta_{S_{i}}}f(\bx | \hat{\btheta}_{S}, \hat{\bbeta}_{0})}^2\right)_{1\leq i \leq L}.$$

To derive the prior induced on the scales, we first need to compute $\kl{p}{q}$. 
Now that we have two normal models, we can get an analytic expression for the \rm{KL} distance recall that the \rm{KL} divergence between two normal distributions is:
:
\begin{align}
    2\kl{ p\left( f(\bx)| \tildetau, \bc \right)}{p_{0}} = & \frac{\norm{\nabla_{\hat{\bbeta_{0}}}f(\bx | \hat{\btheta}_S, \hat{\bbeta}_{0})}^2 + \tildetau \bc \odot \alpha(\bx)}{\norm{\nabla_{\hat{\bbeta_{0}}}f(\bx | \hat{\btheta}_S, \hat{\bbeta}_{0})}^2} - \log\left(\frac{\norm{\nabla_{\hat{\bbeta_{0}}}f(\bx | \hat{\btheta}_S, \hat{\bbeta}_{0})}^2 + \tildetau \bc \odot \alpha(\bx)}{\norm{\nabla_{\hat{\bbeta_{0}}}f(\bx | \hat{\btheta}_S, \hat{\bbeta}_{0})}^2}\right) - 1 \nonumber \\ 
             = & \tildetau \bc \odot \frac{\alpha(\bx)}{\norm{\nabla_{\hat{\bbeta_{0}}}f(\bx | \hat{\btheta}_S, \hat{\bbeta}_{0})}^2} - \log\left( 1 + \tildetau \bc \odot \frac{\alpha(\bx)}{\norm{\nabla_{\hat{\bbeta_{0}}}f(\bx | \hat{\btheta}_S, \hat{\bbeta}_{0})}^2}\right) \nonumber \\
             = &  \tildetau \tildealpha_{\bc}(\bx) - \log\left( 1 + \tildetau \tildealpha_{\bc}(\bx)\right)  \mbox{ where } \tildealpha_{\bc}(\bx) = \frac{\bc \odot \alpha(\bx)}{\norm{\nabla_{\hat{\bbeta_{0}}}f(\bx | \hat{\btheta}_S, \hat{\bbeta}_{0})}^2}\nonumber \\
             =&  \tildetau\tildealpha_{\bc}(\bx) \mbox{ with } \phi(x) = x - \log(1+x).
\end{align}

\subsection*{Checking the approximation}\label{Approx_check}
The complete procedure to evaluate the how the approximation in \eqref{eq: first_order} will result in the appropriate distribution on the scaled KL divergence is detailed in the following steps:
\begin{itemize}
    \item Let the architecture of the network be $\{64, 32, 1\}$
    \item Generate $K=50$ target data sets $\bX_k$ (no need for the output);
    \item For each data set generate random parameters using the Glorot method \citep{Glorot10understandingthe} and do the following.
\begin{itemize}
    
    \item Compute the gradient of model output for each entry in each data set and specify $\lambda$ value for $\btau$ distribution;
    \item Sample $S=20000 \, \btau$'s; and for each sampled $\btau$ do:
    \begin{enumerate}
        \item Sample $M=1000$ parameters $\btheta_i|\widehat{\btheta}, \btau$ and compute corresponding NN output for each target data set $\bX_k$;
        \item Now we have $N*S$ outputs from model $f_{\btheta_i}$ for each target $\bX_k$;
        \item We want to approximate for each $k = 1 \ldots K$:
    $$ KL(p||q) = \mathbf{E}\left[ \log \left( \frac{p(Y)}{q(Y)} \right) \right] \mbox{ where }  \quad Y \sim p $$
    so we use Monte Carlo integration,
    $$ \widehat{KL}(p||q) = N^{-1}\sum_{i =1} ^{ N} \log \left( \frac{p(f(X_i))}{q(f(X_i))} \right)  \mbox{ where } \quad f(X_i) \iid p() ;$$
        \item We also need to approximate $p(Y_i)$ using MC integration, 
    $$ \log\left(p(f(X_i))\right) \approx  \log\left[M^{-1}\sum_{j = 1}^{M} \mathcal{N}(f(X_i); f(X_i; \btheta_j), \sigma)\right]$$ 
        \item Compute the KL-divergence empirically using steps 3 and 4;
    \end{enumerate}
    
    \item Now we have a $\widehat{\rm{KL}}$ value for each sample $\btau$;
    \item Plot the empirical cumulative survival function against theoretical one for pre-specified value of $\lambda$.
\end{itemize}
\end{itemize}

\section*{Appendix A.2: Optimization Details}\label{s:A2}

For the training of all the non-Bayesian models we used stochastic gradient optimization \citep{bottou2012stochastic} with the Adam optimizer \citep{kingma2014adam} implemented in {\tt Tensorflow} and {\tt Keras} python libraries. We use default tuning parameters, e.g., the learning rate is initialized at $0.001$ and decreased by a factor of $0.7$ when the loss function reaches a plateau. Furthermore, a combination  of  L1 and L2 regularization for the parameters of the models. The choice of the batch size was performed using cross-validation for all data sets and models. We considered $\{16, 32, 64, 128\}$ possible choices. Furthermore, to avoid overfitting early stopping was used. Training is terminated when the validation loss does not decrease for $6$ successive epochs.

The following hyperparameters were used for MegNet: the number of blocks was $4$, the embedding dimension was $32$, the L2 coefficient was $0$, the $r$ cutoff was $4$ and $\mbox{nfeat}\_\mbox{bond}$ was $150$.  The following hyperparameters were used for SchNet: n\_atom\_basis was $127$, \n\_filters was $32$, \n\_interactions was $2$, cutoff was $5$ and n\_gaussians was $25$.
    
\section*{Appendix A.3: MCMC Details}\label{s:A3}

\begin{figure}[H]
    \centering
    \includegraphics[scale=.175, trim={110mm 100mm 100mm 100mm}, clip]{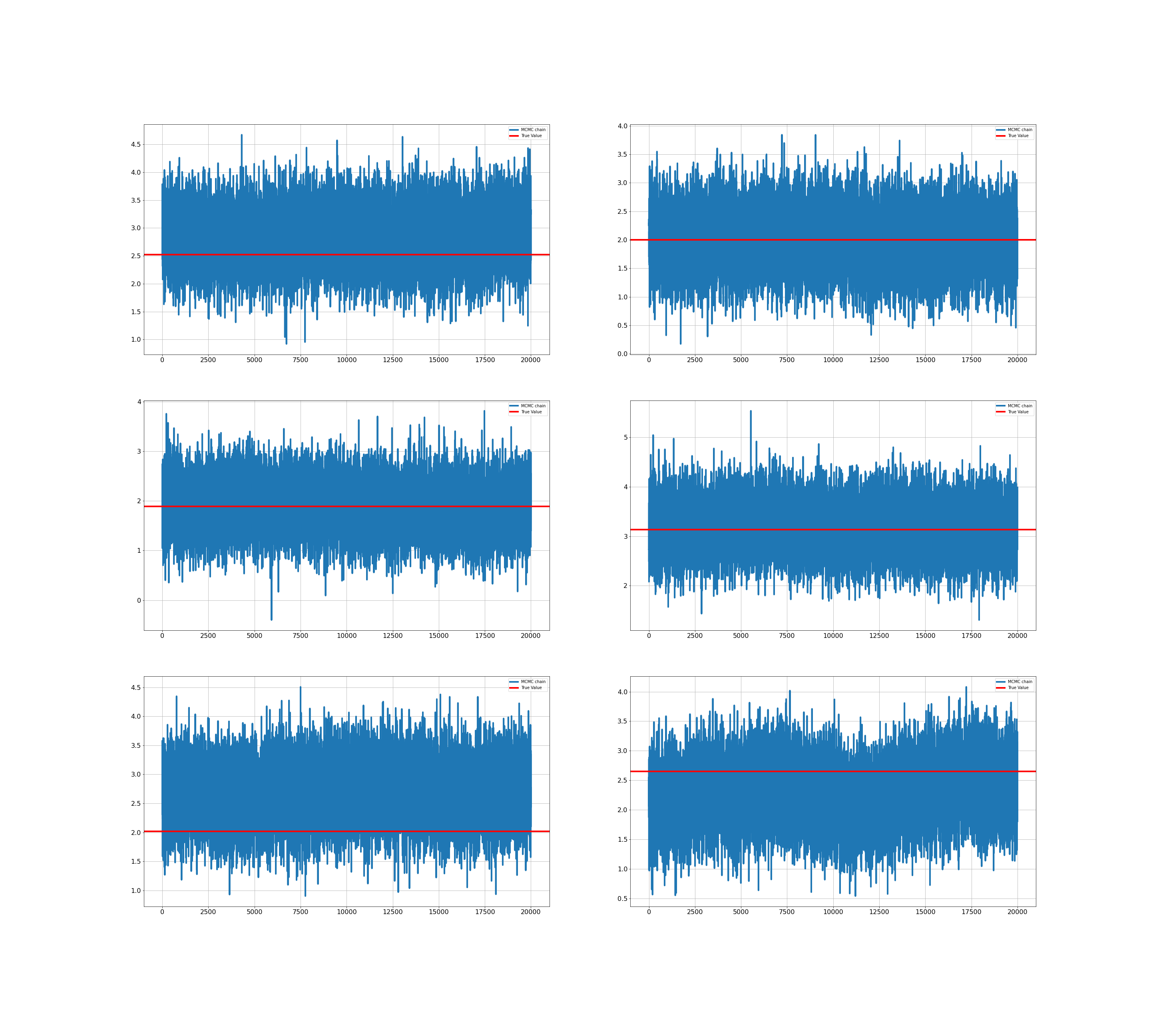}
    \caption{MCMC evolution for six randomly drawn fitted values $f(\bx_i; \btheta)$ of the training set.}
    \label{fig:mcmc_chain}
\end{figure}

All the Bayesian methods in this work were fit using Hamiltonian Monte Carlo \citep{brooks2011handbook} method to sample the posterior distribution of all the parameters using the {\tt Tensorflow-Probability} library in Python.  For the simulation study, the step size was tuned to approximately get an acceptance probability of $0.65$, while the number of leapfrog steps was fixed at $50$. The number of burn-in samples for every chain was $100,000$ and $10,000$ samples were collected with a thinning of $4$.  For the case study of Band Gap prediction, we used the No-UTurn-Sampler \citep{hoffman2014no} to automatically determine the step size and the integration time. Twenty preliminary runs of $5000$ samples each were first used to estimate the mass matrix, then $300,000$ samples were discarded as burn-in and $20,000$ samples were collected after thinning with $10$ samples. Figure \ref{fig:mcmc_chain} shows the evolution of the Markov chain during the sampling stage.

\bibliographystyle{imsart-nameyear}
\bibliography{refs}
\end{document}